# Variational principle for shape memory alloys

Vladimir Grachev, Yuriy Neustadt[a]

Samara State Technical University (SamGTU), Samara, Russia

[a]neustadt99@mail.ru

## Abstract

The quasistatic problem of shape memory alloys is reviewed within the phenomenological mechanics of solids without microphysics analysis. The assumption is that the temperature variation rate is small. Reissner's type of generalized variational principle is presented, and its mathematical justification is given for three-dimensional bodies made of shape memory materials.

## Introduction

Alloys like nitinol, the mechanics of which has been intensively studied for decades (Funakubo (Ed.), 1987; Brinson et al., 1996; Savi et al., 2002; Auricchio et al., 2007; Lagoudas, (Ed.), 2008), differ from many other metals because they can restore their initial shape after plastic deformation (force action) and heating. The stress-strain state of solids consisting of shape memory alloys is analyzed based on physical models, where martensitic and austenitic phase transformations occur under external forces and temperatures (Niezgodka et al., 1988; Bénilan et al., 1990; Auricchio and Sacco, 1997; Bonetti et al., 2016). Relevant mathematical problems are being studied with the spaces of generalized functions using variational inequalities. Many interesting problems have been solved toward this goal.

It should be noted that shape memory materials have one distinctive characteristic: the shape of the irreversibly changed specimens completely restores after loading and unloading (at constant temperature) if the specimens are heated at certain temperature (for each alloy). In other words, phenomenologically "initial and reverse stresses" occur the same way, up to a sign. Because force loading causes plastic deformation, this brings up the question of whether it is possible to explain "reverse stress" based on the theory of plasticity without martensitic and austenitic transformations. The proposed paper serves this purpose. The variational principle of Reissner's type is presented and proved for the existence of the Lagrangian function saddle point defined for generalized velocity of strain and stress set in four-dimensional space and time. The developed variation principle by Reissner (Reissner, 1950, 1965) combines two main local minimum principles within the theory of perfect plasticity: minimum principle for strain rates and minimum principle for field of stress velocities (Koiter, 1960).

The statement of variational principles covers heuristic considerations on the definition of variational principles for shape memory materials without specific identification of input functional spaces. The next section contains mathematical definitions and proof of the principles. The main difficulties in proving the theorem of existence, based on the variational principle, are clearly seen on the idealized model when there is no hardening and the smooth loading surface is replaced with the von Mises surface. This simplified ideal plastic model is studied later, and consideration of translational hardening does not cause any further complications.

## Statement of variational principles

In the case of monoaxial tension of ideal shape memory plastic materials at a constant temperature, the stress-strain curves depend on the temperature and look similar to those illustrated in Figure 1, $T_0 < T_1 < T_{2\varepsilon}$.

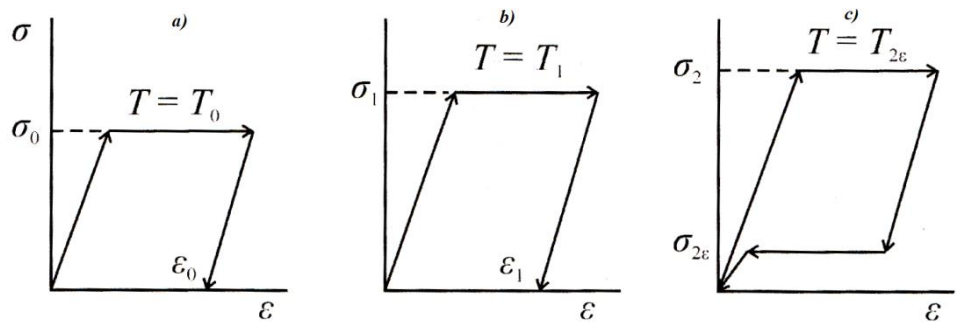

**Figure 1.** Stress-strain curves for ideal shape memory material

Deformation at "room" temperatures $T_0, T_1$ corresponds to behavior of ideal elastic-plastic materials with yield strength $\sigma_0, \sigma_1$. If the tests are done at temperature $T_{2\varepsilon}$, then two surfaces $\sigma_1$ and $\sigma_{2\varepsilon}$ are observed, where large deformations of different signs are traced. There is no residual deformation in the last case. If $\sigma_{2\varepsilon} \to 0$, then $T_{2\varepsilon} \to T_2$, and the value $T_2$ is called the shape recovery temperature.

The curves in Figure 1 allow for another interpretation. Let us consider two stages of medium deformation. At stage one, we would apply stress $\sigma \leq \sigma_0$ at temperature $T_0$. Then, we would relieve the load and measure residual strain $\varepsilon_0$. At stage two, we would raise the temperature to $T_2$ without extra stress. The specimen should return to its initial state.

Let us mathematically demonstrate setting of the deformation problem at stage one.

We would study the problem in relation to a solid three-dimensional statement with the von Mises yield criterion.

$$s_{ij}^2 = 2k^2; \quad s_{ij} = \sigma_{ij} - \sigma\delta_{ij}, \quad \sigma = \sigma_{ii}/3 \tag{1}$$

Here, $k$ is the shear yield strength of the material, $s_{ij}$ is the stress deviator, $\sigma_{ij}$ is the stress tensor, and $\delta_{ij}$ is Kronecker's delta.

Following (Koiter, 1960), we would call stress tensor $\sigma_{ij}$ acceptable if its deviator met the formula $s_{ij}^2 \leq 2k^2$. If deviator $s_{ij}^0$ met inequation $s_{ij}^0 s_{ij}^0 \leq 2k_0^2$, $k_0 < k$, then we would call the corresponding tensor $\sigma_{ij}^0$ safe.

For the parameters defining the continuum behavior, we choose stress tensor $\sigma_{ij}$ and scalar $\lambda$ correlating with the plastic strain rate tensor $\varepsilon_{ij}^p$ as follows (Koiter, 1960; Kachanov, 1971):

$$\dot{\lambda} = \lambda_{2ij}\varepsilon_{ij}^p; \quad \varepsilon_{ij}^p = \varepsilon_{ij} - \varepsilon_{ij}^e, \quad \varepsilon_{ij}^e = E_{ijkl}^{-1}\dot{\sigma}_{kl} \tag{2}$$

The point designates time differentiation, $\varepsilon_{ij}$ is the strain rate tensor, and $E_{ijkl}$ is the elastic modulus tensor. The tensor $\lambda_{2ij}$ is taken per Prager's recommendation (Prager, 1958):

$$\lambda_{2ij} = \sigma_{ij} \tag{3}$$

The internal energy density variation rate is calculated per formula

$$\dot{U} = E_{ijkl}^{-1}\sigma_{ij}\dot{\sigma}_{kl} + \dot{\lambda} \tag{4}$$

based on the following requirement: the process of elastic-plastic straining of solids at constant temperature and with no heat inflow should be described by Prandtl-Reuss law.

$$\varepsilon_{ij} = \varepsilon_{ij}^e + \varepsilon_{ij}^p = E_{ijkl}^{-1}\dot{\sigma}_{kl} + \varepsilon_{ij}^p$$

The formal contracting of this expression with tensor $\sigma_{ij}$ leads to equality,

$$\dot{U} = \sigma_{ij}\varepsilon_{ij}, \quad \dot{U} = E_{ijkl}^{-1}\sigma_{ij}\dot{\sigma}_{kl} + \sigma_{ij}\varepsilon_{ij}^p$$

and integration of the last expression by volume and time matches the first law of thermodynamics when there is no heat inflow ($\dot{q} = 0$).

Thus, if the Prandtl-Reuss law is taken as the defining formula, then the variation rate of internal energy density is calculated per (4), where $\dot{\lambda} = \sigma_{ij}\varepsilon_{ij}^p$. Physically, $\lambda$ is the power necessary for plastic deformation of the volume unit.

The same result would be obtained if we formally required satisfaction of equations (2) - (4) and removed the rule of separating elastic deformation from plastic deformation within the laws of thermodynamics. Indeed, at $\dot{q} = 0$, the first law of thermodynamics means

$$\dot{U} = \sigma_{ij}\varepsilon_{ij} = E_{ijkl}^{-1}\sigma_{ij}\dot{\sigma}_{kl} + \sigma_{ij}\varepsilon_{ij}^p \tag{5}$$

which is equivalent to the division of deformation into elastic and plastic parts:

$$\varepsilon_{ij} = \varepsilon_{ij}^e + \varepsilon_{ij}^p = E_{ijkl}^{-1}\dot{\sigma}_{kl} + \varepsilon_{ij}^p \tag{6}$$

Finally, we use Drucker's postulate on normality of the tensor $\varepsilon_{ij}^p$ to the loading surface (1) and find

$$\varepsilon_{ij}^p = \lambda_p s_{ij}, \quad \varepsilon_{ii}^p = 0 \tag{7}$$

Substituting expression (7) to condition (1), we define parameter $\lambda_p$ and deviator $s_{ij}$

$$\lambda_p^2 = \varepsilon_{ij}^p \varepsilon_{ij}^p / 2k^2, \quad s_{ij} = k\sqrt{2}\varepsilon_{ij}^p (\varepsilon_{ij}^p \varepsilon_{ij}^p)^{-1/2} \tag{8}$$

The formula for the internal energy density variation rate transforms as follows:

$$\dot{U} = f + h, \quad f = E_{ijkl}^{-1}\sigma_{ij}\dot{\sigma}_{kl}, \quad h = k\sqrt{2}(\varepsilon_{ij}^p \varepsilon_{ij}^p)^{1/2} \tag{9}$$

Thus, stating equations (3) and (4), we find division of deformation into reversible and irreversible parts and automatic application of the first law of thermodynamics under isothermal deformation of solids without heat inflow. The second law of thermodynamics also applies automatically because equation $f$ can be identified with the change rate of the Helmholtz free energy, while no heat inflow $\dot{q} = 0$ and formula (9) give the following expression:

$$0 = \dot{q} < T\dot{s} = h$$

where $T$ is the temperature, and $\dot{s}$ is the entropy change rate that combines in one formula two main laws of thermodynamics under isothermal loading.

As a result, the set problem of isothermal deformation of elastic-plastic solids includes neither temperature nor laws of thermodynamics in its final form. It is still unclear how "latent heat of plastic melting" $h$ transforms. This portion of energy is assumed to be scattered into the environment. The last process runs so fast (or deformation is so slow) that the main flow parameters do not change. These are the processes that are studied further.

Born and Furth (Furth, 1940) noticed similarities between plastic yielding at constant temperature and melting. Both phenomena are followed by scattering of internal (latent) energy into the environment. Formula (4), like the Prandtl-Reuss equations, ensures a certain mechanism of scattering: in the beginning, external force energy at a constant temperature converts into internal energy of the specimen, then the latter one scatters into the atmosphere with a heat conductivity factor equal to infinity.

Let the system of volume forces $X_i$ act on the elastic-plastic body that occupies the area $D$ with the boundary $\partial D = \partial D_u + \partial D_p$. On surface $\partial D_u$, the velocities are equal to zero, and on $\partial D_p$, the surface forces are equal to zero. The time interval when straining occurs is designated [0, t]. The assumption is that at any time there is a safe, statistically acceptable distribution of stresses $\sigma_{ij}^0$, when the following equality is valid at any velocities $u_i$,

$$\int_D \sigma_{ij}^0 \varepsilon_{ij} dx - \int_D X_i u_i dx = 0 \tag{10}$$

The problem of elastic-plastic behavior of solids is to find such tensor $\sigma_{ij}$ and vector $u_i$, that formula (5) and equations (1) and (7) are satisfied for the function of phase $\dot{U}(\sigma_{ij}, \varepsilon_{ij}^p)$ from (9) for any $\sigma_{ij}$. The inside parameter $\lambda$ is defined from equalities (2) and (3), while the loading vector $X_i$ meets equation (10).

Relying on the results (Mosolov and Myasnikov, 1981; Panagiotopoulos, 1985), it was demonstrated (Neustadt, 1993) that the set problem is equivalent to finding the Lagrangian function saddle point

$$L(\sigma_{ij}, u_i) = \frac{1}{2} \int_{D \times [0,t]} E_{ijkl}^{-1} \dot{\sigma}_{ij} \dot{\sigma}_{kl} dV + \int_{D \times [0,t]} (\dot{\sigma}_{ij}^0 \varepsilon_{ij} - \dot{\sigma}_{ij} \varepsilon_{ij}) dV, \ dV = dx_1 dx_2 dx_3 dt \tag{11}$$

on the field of arbitrary velocities $u_i$ and deviators $s_{ij}^2 \leq 2k^2$.

If equation (10) holds true, then there is a generalized solution to the "minimax" problem (11). This fact follows satisfaction of equation (6) and the first law of thermodynamics as (5) (at constant temperature $T = const$). Given that $\dot{\lambda} \geq 0$, the second law of thermodynamics holds true in the form of a Clausius-Duhem inequality

$$\dot{q} \leq T\dot{s} \tag{12}$$

because when there is no external heat inflow ($\dot{q} = 0$), the rate of internal energy variation can be presented as

$$\dot{U} = f + T\dot{s}$$

where $T\dot{s}$ is the energy dissipation rate.

The mechanical meaning of saddle point existence in solids is as follows: there is a solution as long as the loading allows specifying an acceptable tensor $\sigma_{ij}^0$ at any point of the body.

Let us make sure that in case of "reverse" stress (if temperature is constant), the principle of the fixed point value can be used for a functional similar to (11).

We relieve stress $\sigma_{ij}^0$ to $\varepsilon_0 \sigma_{ij}^0$, where $\varepsilon_0$ is a small number. The stress state at point $x_i$, which is plastically strained at loading $\sigma_{ij}^0$, corresponds to stress tensor A in space and tensor B – at loading $\varepsilon_0 \sigma_{ij}^0$ (Figure 2).

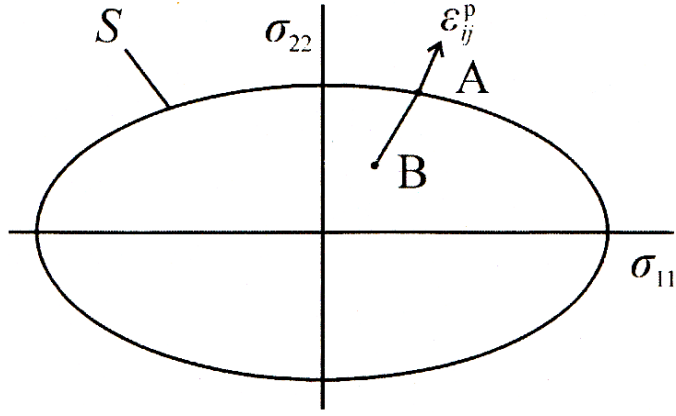

**Figure 2.** Loading surface for ideal elasto-plastic material

The solid is heated to temperature $T_{2\varepsilon}$ (Figure 1c). Let us look at the nine-dimensional manifold $R$, limited with surfaces $\{S : s_{ij}^2 = 2k_2^2\}$ and $\{S_\varepsilon : s_{ij}^2 = 2\varepsilon^2 k_2^2\}$. The surface $S_\varepsilon$ is a result of similar transformation (dilation) of the surface $S$ with the small number $\varepsilon$. The tensor $\sigma_{ij}$ is considered acceptable if the inclusion $\sigma_{ij} \subset R$ is satisfied (Figure 3).

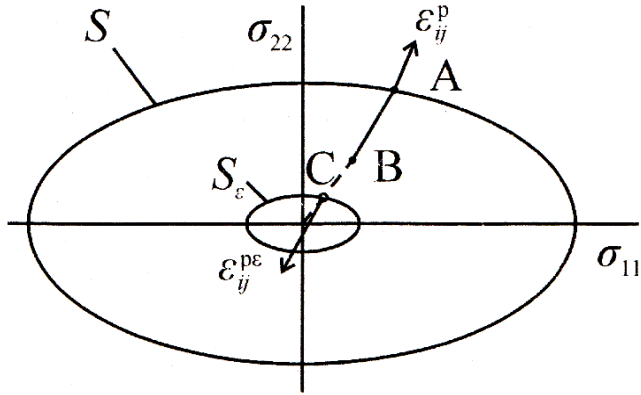

**Figure 3.** Loading surface for ideal elasto-plastic shape memory material

The tensor $\sigma_{ij}$ is considered safe if there is inclusion $\sigma_{ij} \subset R_0$. The manifold $R_0 \subset R$ is limited with surfaces $\{S^0 : s_{ij}^2 = 2k_{02}^2\}$ and $\{S_\varepsilon^0 : s_{ij}^2 = 2\varepsilon_1 k_2^2\}$, $k_{02} < k_2$, $\varepsilon_1 > \varepsilon$.

If we introduce a hypothesis that at temperature $T_{2\varepsilon}$, under stress $\varepsilon_0 \sigma_{ij}^0$, "plastic" strain occurs in alloys like nitinol, requiring extra heat input, while the stresses $\sigma_{ij}$ are acceptable, then it is quite easy to explain the shape memory effect in compliance with the laws of elastic-plastic behavior. We keep supplying heat at the rate $\dot{q}$. If, during this process, point B appears on the surface $S_\varepsilon$, then plastic strain $\varepsilon_{ij}^{p\varepsilon} = -\varepsilon_{ij}^p$ is possible due to similarity of the surfaces $S$ and $S_\varepsilon$. It is clear that the above reasoning is valid if Druker's postulate is not used in its initial form $(\sigma_{ij} - \sigma_{ij}^*)\varepsilon_{ij}^p \geq 0$, where $\sigma_{ij}$ is a real field and $\sigma_{ij}^*$ is every possible stress field, but the local con-clusion $\dot{\sigma}_{ij}\varepsilon_{ij}^p = 0$ (or Ziegler's conditions on orthogonality of thermodynamic forces and flows) is used instead. This approach is often used in problems of thermal plasticity (Ranieck and Sawczuk, 1975).

Let us choose the stress tensor $\sigma_{ij}$ and energy $\lambda$ necessary for "reverse transformation" as internal parameters of the solid. In equation (2) we use the tensor $\lambda_{2ij} = \sigma_{ij}^A$ similar to Figure 1, where $\sigma_{ij}^A$ is a stress tensor at point A (Fig. 3). In other words, "reverse" flow on the surface $S_\varepsilon$ requires as much energy as flow on surface $S$. On the surface $S_\varepsilon$ we have equation (9) by replacement of $h$ with

$$h_2 = k_2\sqrt{2}(\varepsilon_{ij}^p \varepsilon_{ij}^p)^{1/2} \qquad (13)$$

where parameter $\lambda$ is replaced with $h_2$ in equation (4), and in equation (8), number $k$ becomes equal to $\varepsilon\, k_2$. The first law of thermodynamics transforms into

$$\dot{q} = \dot{U} - \sigma_{ij}\varepsilon_{ij} = E_{ijkl}^{-1}\sigma_{ij}\sigma_{kl} - \sigma_{ij}\varepsilon_{ij} + h_2 \qquad (14)$$

Equations (10) and (12) remain the same.

Thus, we have the following similarity: force deformation of shape memory materials (when $\dot{q}=0$) and "reverse thermal restoration" (without forces) only differ in sign change $\varepsilon_{ii}^p$ in the mechanics of solids. Phenomenologically, the process runs the same in both cases: the power of plastic deformation $\lambda$ depends only on $\sigma_{ij}$ and $\varepsilon_{ii}^p$, while factor $k$ changes in the equation for internal energy (9) or (13). Under reverse thermal deformation, the role of the "latent heat of plastic melting" $h$ is played by the "latent heat of reverse transformation" $h_2$, and the external forces power $\sigma_{ij}\varepsilon_{ij}$ is replaced with the heat input rate $\dot{q}$ in the first law of thermodynamics ("under reverse deformation without forces").

Therefore, at the stage of thermal deformation (at constant temperature), it is necessary to find such a stress variation rate tensor $\dot{\sigma}_{ij}$ and a velocity vector $u_i$ that equation (14) is satisfied on the acceptable set of deviators $s_{ij}$ and the saddle point of the Lagrangian function exists where the value $\sigma_{ij}^0$ is replaced with $\varepsilon_0\sigma_{ij}^0$.

The variational principle in this form is proved in Neustadt, 2008. The next section covers generalizations for the problems when temperature and external stress change during loading. The equation for the internal energy variation rate can be used as (9) (with the factor $k(T)$ depending on temperature Bertram, 1982). Only equations of the first and second laws of thermodynamics (10), (12) are adjusted. It is also important that the temperature changes quite slowly, and sup-plied heat is used selectively (differentially), either for "shape restoration" or for temperature change. Shape memory alloys must absorb heat, i.e., release incoming energy into the environment at those points where the stress state corresponds to plastic flow.

**Validation of the variational principle under changing temperature and stress**

Let the solid occupy the three-dimensional area $D$ with the boundary $\partial D = \partial D_u + \partial D_p$. The velocity vector $u_i$ is set on a part of the boundary $\partial D_u$, while the stress rate tensor $\dot\sigma_{ij}$ is known on $\partial D_p$. We will study the problem on the time interval $[0, t]$. At first, we define the four-dimensional area $M = D \times [0, t]$ with the boundary

$$\partial M = \partial M_u + \partial M_p, \quad \partial M_u = \partial D_u \times [0, t], \quad \partial M_p = \partial D_p \times [0, t]$$

Then, we introduce a Hilbert space of functions $\dot H$ as completion of tensors $\sigma_{ij}$ differentiated with respect to time in the norm determined by the scalar product

$$(\dot\sigma_{ij}^1, \dot\sigma_{ij}^2) = \int_M \dot\sigma_{ij}^1 \dot\sigma_{ij}^2 dm, \quad dm = dxdt, \quad dx = dx_1 dx_2 dx_3 \tag{15}$$

and a space of possible velocities as a set of measures with the norm

$$\|u_i\|_{BD(M)} = \max_{ij} \{ \int_{M^+} \varepsilon_{ij}(u_i, dm) + \int_{M^-} | \varepsilon_{ij}(u_i, dm) | \} \tag{16}$$

where $M = M^+ + M^-$ is specified, while sub-areas $M^+$ and $M^-$ divide the area $M$ in such a way that the sign of $\varepsilon_{ij}$ is positive inside of $M^+$ and negative inside of $M^-$.

The space of bounded deformations $BD(M)$ is obtained by completion of the set of functions

$$2\varepsilon_{ij}(u_i, dm) = (\partial u_i / \partial x_j + \partial u_j / \partial x_i) dm$$

in the norm (16). This space is not reflexive. It was studied in relation to viscoplasticity problems (Panagiotopoulos, 1985). Particularly, it was proved that the functions from $BD(M)$ have a trace on the piecewise-smooth surface $\partial D_u$ that belongs to the space of integrable functions. It allows specifying the following constraint for the fixed boundary $\partial M_u$:

$$\|u_i\|_{L_1(M_u)} = 0 \tag{17}$$

Lack of stresses on $\partial M_p$ will be written as follows

$$\|\sigma_{ij}(\partial M_p)\|_{L^\infty(M)} = 0 \tag{18}$$

Here, $L^\infty(M)$ is a space of constrained measurable functions on the set $M$.

The convex sets $\sigma_{ij}$ and $u_i$ are chosen within the spaces $\dot{H}$ and $BD(M)$ so that conditions (17) and (18) and following equations are satisfied (hereinafter, unless otherwise stated, integration is done on area $M$):

$$R = \{\sigma_{ij} : \sigma_{ij} \in L^\infty(M), \quad 2\varepsilon^2 k_2^2 \leq s_{ij}^2 \leq 2k_2^2, \quad s_{ij} = \sigma_{ij} - \sigma_{kk}\delta_{ij}/3 \} \quad (19)$$

$$Q = \int (q(t) + q'(t, \sigma_{ij}))dm = \int (E_{ijkl}^{-1}\sigma_{ij}\sigma_{kl} - \sigma_{ij}\varepsilon_{ij} + k_2(T)\sqrt{2}(\varepsilon_{ij}^p \varepsilon_{ij}^p)^{1/2})dm \quad (20)$$

$$\int T_{2\varepsilon}\dot{s}dm \leq \int (q(t) + q'(t, \sigma_{ij}))dm, \quad \lambda = \int k_2(T)\sqrt{2}(\varepsilon_{ij}^p \varepsilon_{ij}^p)^{1/2}dt = Ts, \quad T\dot{s} = \dot{\lambda} - \dot{T}\lambda/T \quad (21)$$

Condition (19) means that only acceptable stress states (Figure 3) are considered, equations (20) and (21) are the first and second laws of thermodynamics, and the function $k_2(T)$ depends on the temperature. Let us highlight that the laws of thermodynamics "control" plastic deformations while the temperature is assumed to be a known function of time. In formulas (20) and (21), $\dot{q}(t)$ represents the rate of heat input, and $\dot{q}'(t)$ is the heat absorption rate. Herewith,

$$\dot{q}' = 0, \text{ when } \sigma_{ij} \in S_\varepsilon; \text{ in other cases } \dot{q}' + \dot{q} = 0 \quad (22)$$

Finally, let us apply a load $X_i$ to the solid that could be balanced with a safe stress tensor $\sigma_{ij}^0$:

$$2\varepsilon_1 k_2^2 < s_{ij}^0 s_{ij}^0 < 2k_{02}^2, \quad s_{ij}^0 = \sigma_{ij}^0 - \sigma_{kk}^0 \delta_{ij}/3, \quad k_{02} < k_2, \quad \varepsilon_1 > \varepsilon$$

$$\int_D \sigma_{ij}^0 \varepsilon_{ij} dx - \int_D X_i u_i dx = 0 \quad (23)$$

Now, we will write the Lagrangian function

$$L(\sigma_{ij}, v_i) = \frac{1}{2}\int E_{ijkl}^{-1}\dot{\sigma}_{ij}\dot{\sigma}_{kl} dm + \int (\sigma_{ij}^0 - \sigma_{ij})\varepsilon_{ij} dm \quad (24)$$

and prove the existence of $\sigma_{ij}, u_i$ on the set $R \times BD(M)$ under conditions (17) - (23) that correspond to the saddle point of the Lagrangian function $L(\sigma_{ij}, u_i)$.

The proof involves fulfillment verification of the conditions of the theorem below (Ekland and Temam, 1976): if there is an element $u_{0_i} \in BD(M)$ for the sets with constraints (15) - (23) that

$$\lim L(\sigma_{ij}, u_{0_i}) = \infty, \quad \sigma_{ij} \in R, \quad \|\dot{\sigma}_{ij}\|_{\dot{H}} \to \infty \quad (25)$$

and the following equation is satisfied

$$\lim \inf L(\sigma_{ij}, u_i) = -\infty, \quad u_i \in BD(M), \quad \sigma_{ij} \in R, \quad \|u_i\|_{BD(M)} \to \infty \quad (26)$$

then the functional $L(\sigma_{ij}, u_i)$ has a saddle point on $R \times BD(M)$

$$L(\sigma_{ij}, u_i) = \min_{\sigma_{ij} \in R} \sup_{u_i' \in BD(M)} L(\sigma_{ij}', u_i') = \max_{u_i' \in BD(M)} \inf_{\sigma_{ij}' \in R} L(\sigma_{ij}', u_i') = m_0 \quad (27)$$

and a sequence could be chosen among $\sigma'_{ij}, u'_i$ that $\sigma'_{ij} \to \sigma_{ij}$ is weak in $\dot{H}$, and $u'_i \to u_i$ is weak* in $BD(M)$.

The condition (25) is satisfied when $u_{0i} = 0$, and to check equation (26), we take such tensors $\sigma_{ij} = \sigma^0_{ij} + \alpha p_{ij}$ that $p_{ij} = 1$ or $p_{ij} = -1$ depending on integration over sub-area $M^+$ or $M^-$. Sub-areas $M^+$ or $M^-$ are chosen so that the sign $\sigma^0_{ij}$ in them is positive and negative, respectively. If $\alpha > 0$ is sufficiently small because of equation (23), the inclusion $\sigma_{ij} \in R$ holds true. In virtue of equation (24), we have

$$\inf_{\sigma_{ij} \in R} L(\sigma'_{ij}, u_i) \leq -c \|u_i\|_{BD(M)} + c_1 \tag{28}$$

Here, $c$ and $c_1$ are some invariables. The inequality (28) results in the satisfaction of equation (26), and the mixed variational principle is proved.

It should be noted again that the physical meaning of the constraints (20)-(22) is that the input heat is used either for the temperature change within the body or for irreversible plastic deformations on the surface $S_\varepsilon$. In this case, the energy that corresponds to plastic flow on the surfaces $S$, $S_\varepsilon$ scatters into the environment.

If $\sigma_{ij}$ and $u_i$ are functions differentiated on coordinates and time, then we can replace the area $M$ with $D$ in equation (24) and realize the validity of the Prandtl-Reuss law. Indeed, the variation $L(\sigma_{ij}, u_i)$ is presented as

$$\int_D (E^{-1}_{ijkl} \dot{\sigma}_{kl} - \varepsilon_{ij}(u_i))(\sigma_{ij} - \sigma'_{ij}) dx \geq 0 \tag{29}$$

Due to the arbitrariness of $(\sigma_{ij} - \sigma'_{ij})$ everywhere except of the points of the surface,

$$\Phi = s^2_{ij} - 2k^2_2 \varepsilon^2 = 0 \tag{30}$$

we can write the following equation:

$$\varepsilon_{ij} = E^{-1}_{ijkl} \dot{\sigma}_{kl} + \lambda \partial \Phi / \partial s_{ij}; \tag{31}$$

this is the Prandtl-Reuss law, which follows that for the differentiable functions satisfying the flow on the surface (30), equation (20) becomes

$$\int q dm = \int ((1-\varepsilon) k_2 \sqrt{2} (\varepsilon^p_{ij} \varepsilon^p_{ij})^{1/2}) dm$$

The integral in the right part of the formula is non-negative; therefore, the heat input is directed, and the second law of thermodynamics is an effect of the first one.

Equation (30) raises the possibility of replacement of the surface (30) with any other smooth surface $\Phi(s_{ij}) = 0$ under yielding conditions. In formula (8), $s_{ij}$ should be replaced with $\partial\Phi/\partial s_{ij}$, and the expression $(\varepsilon_{ij}^p \varepsilon_{ij}^p)^{1/2}$ should be replaced with the dissipative potential that is a transformation of the Legendre function $\Phi$.

Let us review the sequence $\sigma_{ij}^\varepsilon, u_i^\varepsilon$ when $\varepsilon \to 0$ is under condition (19). Because the value $\max \|\sigma_{ij}^\varepsilon\|$ is limited in view of condition (19), the integral is limited at any $\varepsilon$:

$$U_p = \int (\varepsilon_{ij}^p \varepsilon_{ij}^p)^{1/2} dm$$

Therefore, when $\varepsilon \to 0$ (Josida, 1965), a subsequence can be chosen from the sequence $u_i^\varepsilon$ that weakly* tends to a limit within $BD(M)$. It could be taken as a solution to the problem of the "reverse flow" of the shape memory alloys at temperature $T_2$.

It is clear that the variational principle does not hold true in the limiting state because two first members of the subintegral equation "disappear" in the right part of formula (20) in the most important special case when $\sigma_{ij} \to 0$. The first law of thermodynamics transforms into the obvious equality that describes one-dimensional flow because only the second invariant of the tensor $\varepsilon_{ij}^p$ is "remembered".

It should also be noted that it is required to prove regularity of the tensor $\sigma_{ij}$ and the vector $u_i$ in order to convert the extremality condition of the Lagrangian function into equation (29). This problem has been studied by many authors, but is still far from being solved (Lewy and Stampacchia, 1970; Capogna and Garofalo, 2003; Focardi et al., 2017).

**Conclusions**

The key outcome of this research is that the mixed variational principle of Reissner's type can be applied to solids made of shape memory materials if the temperature variation rate is quite small: the process must be both quasi-static and "quasi-isometric".

Phenomenologically, the process can be described by the classical theory of plasticity without physics that takes into account austenitic and martensitic transformations. The further progress of the theory requires a number of mathematical problems to be solved.

In reality, the temperature and heat input rate are not independent. The last constraint can be dropped if the first law of thermodynamics (20) is replaced with the generalized equation of heat conductivity

$$\frac{\partial T}{\partial t} = c_1 \left( \frac{\partial^2 T}{\partial x_1^2} + \frac{\partial^2 T}{\partial x_2^2} + \frac{\partial^2 T}{\partial x_3^2} \right) + c_3 \dot{q}(t) - \Lambda, \quad \Lambda = c_2 (E_{ijkl}^{-1} \sigma_{ij} \sigma_{kl} - \sigma_{ij} \varepsilon + k_2(T) \sqrt{2} (\varepsilon_{ij}^p \varepsilon_{ij}^p)^{1/2})$$

or its solution that is obtained using formulas for the exponent and $E_v$ - spectral decomposition of the Laplace distribution unit (Josida, 1965)

$$T(t,x) = \exp((tB)T(0,x) + \int_0^t \exp((t-s)B) \bullet (c_3\dot{q}(t) - \dot{\Lambda})ds$$

$$\exp((t-x)B) = \sum_{k=0}^{\infty}((t-x)B)^k/k! \ , \ B = \frac{\partial^2 T}{\partial x_1^2} + \frac{\partial^2 T}{\partial x_2^2} + \frac{\partial^2 T}{\partial x_3^2} \ , \qquad B^k = \int_0^{\infty} v^k dE_v$$

In addition, we need to know the initial temperature distribution $T(0, x)$ inside the solid and the fixed surface temperature $\partial D$ the entire time. Letters $c_1, c_2, c_3$ designate some constants. The constant $c_2$ is not equal to zero only at those points of the solid where stress is located on the loading surface $S_\varepsilon$. The strong evidence of the given assumption for various classes of generalized functions (Ladyzhenskaya et al., 1968; Evans, 2010) seems interesting and important. Other interesting problems include improvement of computational procedures and the regularity assessment of solutions following the mixed variational principle.

# References

Auricchio F and Sacco E (1997) A Superelastic Shape-Memory-Alloy Beam Mode. *Journal of Intelligent Material Systems and Structures* 8. Issue 6: 489-501.
Auricchio F, Reali A and Stefanelli U (2007) A three-dimensional model describing stress-induced solid phase transformation with permanent inelasticity. *Int. J.Plasticity* 23: 207–226.
Bénilan Ph, Blanchard D and Ghidouche H (1990) On a non-linear system for shape memory alloys. *Continuum Mechanics and Thermodynamics* 2, Issue 1: 65–76.
Bertram A (1982) Thermo-mechanical constitutive equations for the description of shape memory effects in alloys. *Nuclear engineering and design* 74: 174-182.
Bonetti E, Colli P, Fabrizio M, et al. (2016) Existence of Solutions for a Mathematical Model Related to Solid–Solid Phase Transitions in Shape Memory Alloys. *Archive for Rational Mechanics and Analysis* 219, Issue 1: 203–254.
Brinson LC, Bekker A and Hwang S (1996) Deformation of Shape Memory Alloy due to Thermo-induced Transformation, *Journal of Intelligent Material Systems and Structures* 7: 97-107.
Capogna L and Garofalo N (2003) Regularity of minimizers of the calculus of variations. *J. Eur. Math. Soc.* 5: 1–40.
Ekland I and Temam R (1976) *Convex Analysis and Variational Problems*. Amsterdam: North-Holland.
Evans LC (2010) *Partial differential equations.* 2nd ed. New York: American Math. Society.
Focardi F, Geraci M and Spadaro E (2017) The classical obstacle problem for nonlinear variational energies. *Nonlinear Analysis: Theory, Methods & Applications* 154: 71-87.
Funakubo H (Ed.) (1987) *Shape Memory Alloys*. New York: Gordon and Breach Science Publishers.
Furth R (1940) Relation between breaking and melting. *Nature* 145, No. 3680: 741–761.
Josida K (1965) *Functional Analysis*. Berlin: Springer.
Kachanov LM (1971) *Foundations of the Theory of Plasticity*. Amsterdam/London: North-Holland Publishing Company.
Koiter WT (1960), General theorems for elastic-plastic solids. In: *Progress in solid mechanics*. Amsterdam, North-Holland publishing company, pp.167-221.
Ladyzhenskaya OA, Solonnikov VA and Uraltseva NN (1968) *Linear and Quasilinear Equations of Parabolic Tipe*. New York: American Math. Society.
Lagoudas DC (Ed.) (2008) *Shape Memory Alloys. Modeling and Engineering Applications*. New York: Springer Science+Business Media.

Lewy H and Stampacchia G (1970) On the smoothness of superharmonics which solve a minimum problem. *J. Anal. Math*. 3: 227–236.

Mosolov PP and Myasnikov VP (1981) *Mechanics of rigid plastic solids*. Moscow: Nauka (in Russian).

Neustadt YS (1993) Generalised solutions in flow theory of ideal elastoplastic solids. *Izvestia of RAS. Mechanics of solids* 6: 74-78 (in Russian).

Neustadt YS (2008) Variational principle for shape memory solids. *Vestnik of the South-Ural University. "Mathematics, physics, chemistry"* 22(122): 4-11 (in Russian).

Niezgodka M, Zheng S and Sprekels J (1988) Global solutions to a model of structural phase transitions in shape memory alloys. *Journal of mathematical analysis and applications* 130: 39-54.

Prager W (1958) Non-isothermal plastic deformation. *Proc. Konikl. Nederl. Acad. Wet.* Bd61. No. 3: 176-182.

Panagiotopoulos PD (1985) *Inequality problems in mechanics and applications*. Boston: Birkhauser.

Raniecki B and Sawczuk A (1975) Thermal effects in plasticity. *ZAMM*. 55, Pt1.: 333-341; Pt2.: 363-373.

Reissner E (1950) On a variational theorem in elasticity. *Jouranal of Mathematical Physics* 29: 90-95.

Reissner E (1965) On mixed variational formulations in finite elasticity. *Inernational Jouranal of Solids and Structures* 1: 93-95.

Savi AM, Pavia A, Baeta-Neves PA, et al. (2002) Phenomenological Modeling and Numerical Simulation of Shape Memory Alloys: A Thermo-Plastic-Phase Transformation Coupled Model. *Journal of Intelligent Material Systems and Structures* 13(5): 261-273.